\documentstyle[preprint,prl,aps]{revtex}
\begin{document}
\tightenlines
\draft

\title{Mixed states on neural network with structural learning}
\author{Tomoyuki Kimoto}
\address{
Oita National College of Technology,
1666,Maki,Oita-shi,Oita 870-0152,Japan}

\author{Masato OKADA}
\address{
Brain Science Institute, RIKEN, 2-1 Hirosawa, Wako-shi, Saitama 351-0198, Japan,\\
ERATO Kawato Dynamic Brain Project, 2-2 Hikaridai, Seika-cho, Soraku-gun, Kyoto 619-0288, Japan
}

\date{\today}
\maketitle

%
%

\begin{abstract}
We investigated the properties of mixed states in a sparsely encoded associative memory model with a structural learning method. When mixed states are made of $s$ memory patterns, $s$ types of mixed states, which become equilibrium states of the model, can be generated. To investigate the properties of $s$ types of the mixed states, we analyzed them using the statistical mechanical method. We found that the storage capacity of the memory pattern and the storage capacity of only a particular mixed state diverge at the sparse limit. We also found that the threshold value needed to recall the memory pattern is nearly equal to the threshold value needed to recall the particular mixed state. This means that the memory pattern and the particular mixed state can be made to easily coexist at the sparse limit. The properties of the model obtained by the analysis are also useful for constructing a transform-invariant recognition model.
\end{abstract}

\narrowtext

%
\section{Introduction}

When an associative memory model is made to store memory patterns 
as a result of correlation learning, 
not only the memory patterns but also states generated by mixing of 
arbitrary memory patterns automatically become 
the equilibrium state of the model. 
These states are called the {\it mixed states}, 
and they are not simply a side effect unnecessary for information processing. 
For example, Amari discussed ``concept formation'' 
using the stability of the mixed states \cite{Amari1977}. 
The correlated attractor \cite{Griniasty1993}\cite{Amit1994}, 
which is a model of the Miyashita attractor \cite{Miyashita1988a}, 
is considered to be a mixed state in a broad sense. 

The sparse coding scheme is thought to be used in the brain 
according to some physiological findings \cite{Miyashita1988b} 
and theoretical viewpoints \cite{Tsodyks1988}\cite{Buhmann1989}\cite{Amari1989}\cite{Perez-Vicente1989}\cite{Okada1996}.
Thus, it is necessary to analyze the properties of mixed states 
in the sparsely encoded associative memory model. 
Kimoto and Okada analyzed the mixed states 
in a sparsely encoded associative memory model \cite{Kimoto2001}.
They found that when the $s$ memory patterns are mixed, 
$s$ types of the mixed states can be generated by them 
and all mixed states can become equilibrium states 
by adjusting the threshold value of the model.
Moreover, among $s$ types of the mixed states, 
the performance of OR mixed state 
generated by OR operation through each element of memory patterns
is the highest at the sparse limit for two reasons.
(1) At the sparse limit of firing rate $f \rightarrow 0$, 
the storage capacity of the OR mixed state diverges as $1/|f \log f|$,
and the storage capacities of the other mixed states become 0. 
(2) At the sparse limit of firing rate $f \rightarrow 0$, 
the threshold value needed to recall the memory pattern 
approximately corresponds to the threshold value needed to recall the OR mixed state, 
and is greatly different from the threshold values needed to recall the other mixed states. 
Therefore, 
the memory pattern and the OR mixed state becomes the equilibrium state 
without any readjustment to the threshold value.

Parga and Rolls proposed a transform-invariant recognition model 
based on an associative memory model 
and analyzed the properties of the model using the statistical mechanics 
\cite{Parga1998}. 
Their proposed model uses a modified Hebbian learning method 
that includes cross-correlation terms between various views (memory patterns) of the same object. 
When the strength of the cross-correlation terms is small, 
the equilibrium state of the model becomes the view itself. 
When the strength exceeds a certain threshold, 
the equilibrium state of the model becomes the mixed state,
which is generated by all views of the same object, 
independent of the particular view that was presented as a stimulus. 
Parga and Rolls showed that, 
in the case of strong cross-correlation terms, 
the state of the model is drawn into a peculiar mixed state of the object,
and the model becomes able to recognize the object by observing the drawn mixed state
regardless of the coordinate transformation. 
Their model is very interesting for two reasons. 
(1) Since all views of the same object are related by cross-correlation terms 
and the views of different objects are not related, 
there is a structure in the learning method. 
(2) The equilibrium state changes from the view to the mixed state depending on the strength of the cross-correlation terms.

Elliffe et al. added a sparse coding scheme to the model 
proposed by Parga and Rolls and examined the properties of the mixed state 
by computer simulation \cite{Elliffe1999}. 
Since the sparse coding scheme is thought to be used in the brain, 
their model is very interesting. 
Then the model should be analyzed not only by computer simulation 
but also by the statistical mechanics. 
Moreover, though their model uses $\mbox{sgn}(\cdot)$ as the neuron output function, 
its storage capacity does not diverge at the sparse limit for this output function. 
The advantage of using the sparse coding scheme is thus reduced by half, 
so the output function should be modified.

We have therefore added a sparse cording scheme and a modified output function 
to an associative memory model with a learning method including cross-correlation terms 
and analyzed the properties of the model using the statistical mechanical method.
%
%
\section{Model}

We consider an associative memory model consisting of $N$ neurons 
with output function $\Theta(\cdot)$. We use synchronous dynamics,
\begin{eqnarray}
  x_i^{t+1} &=& \Theta(\sum_{j \ne i}^N J_{ij} x_j^t + h^t),
                \quad \quad i=1,2 \cdots,N, 
                \label{eq-model1:equilibrium} \\
  \Theta(u) &=& \cases{
		1 & u $\ge$ 0 \cr
        	0 & u  $<$  0 \cr
      		} ,
            \label{eq-model1:theta}
\end{eqnarray}
where $x_i^t$ represents the state of the $i$th neuron at discrete time $t$, 
and $J_{ij}$ denotes the synaptic coupling from the $j$th neuron to the $i$th neuron. 
The threshold value $h^t$ is assumed to be independent of the serial number $i$ of a neuron. 
Its concrete value is described later. 
The output function $\Theta(\cdot)$ is assumed to be a step function, 
as shown in Eq. (\ref{eq-model1:theta}). 
Elliffe's model (1999) used the output function as follows,
\begin{equation}
  \mbox{sgn}(u) = \cases{
		+1 & u $\ge$ 0 \cr
        	-1 & u  $<$  0 \cr
      		} ,
	\label{eq-model1:sgn}
\end{equation}
In this function, 
the storage capacity does not grow at the sparse limit \cite{Okada1996}. 

Memory pattern $\mbox{\boldmath $\eta$}^{\mu \nu}$ 
is a vector of $N$ dimensions composed of the elements 0-1, 
$\mu$ stands for the number of group to which the memory pattern belongs, 
and $\nu$ stands for the number of memory patterns in that group. 
Each group includes $s$ memory patterns. 
Each component $\eta^{\mu \nu}_i$ of memory pattern $\mbox{\boldmath $\eta$}^{\mu \nu}$ is independently generated by the firing rate $f$,
\begin{eqnarray}
    && \mbox{Prob}[\eta^{\mu \nu}_i=1] = 1-\mbox{Prob}[\eta^{\mu \nu}_i=0] = f,
    \nonumber \\
    && i=1,2,\cdots,N. 
	\label{equationpattern} 
\end{eqnarray}
A memory pattern with a small firing rate is called a {\it sparse pattern},
and the use of a sparse pattern for the memory pattern is called {\it sparse coding}. 
The synaptic coupling $J_{ij}$ is determined by the following learning method,
\begin{equation}
  J_{ij} = \frac {1}{Nf(1-f)} 
	\sum_{\mu=1}^{\alpha N} \sum_{\nu=1}^{s} \sum_{\nu'=1}^{s}
	(\eta_i^{\mu \nu} - f) B_{\nu \nu'} (\eta_j^{\mu \nu'} - f),
	\label{eq-model1:covariance} 
\end{equation}
where $\alpha N$ is the number of stored groups, 
$\alpha$ is the loading rate, 
and $B_{\nu \nu'}$ is the strength of the cross-correlation terms 
between the memory patterns ($\nu$ and $\nu'$) of the same group.
\begin{equation}
  B_{\nu \nu'} = \delta_{\nu \nu'} + b(1-\delta_{\nu \nu'}).
	\label{eq-model1:cross-colleration}
\end{equation}

Next, we explain ``mixed state''. 
We discuss the mixed state composed of the $s$ memory patterns 
that belong to the same group. 
We explain the generating method of a mixed state 
as an example of the memory patterns 
$\mbox{\boldmath $\eta$}^{1, 1},\mbox{\boldmath $\eta$}^{1, 2},\cdots, \mbox{\boldmath $\eta$}^{1,s},$ belonging to the first group. 
The $i$th element of the mixed state is set to '1' 
if the number of firing state '1' is $k$ or more 
in the $i$th elements $\eta^{1,1}_i,\eta^{1,2}_i,\cdots,\eta^{1,s}_i$ 
of the $s$ memory patterns $\mbox{\boldmath $\eta$}^{1,1},\mbox{\boldmath $\eta$}^{1,2},\cdots,\mbox{\boldmath $\eta$}^{1,s}$.
Otherwise, it is set to $'0'$.
According to this definition, 
the $s$ types of mixed states exist because $1 \leq k \leq s$. 
Even if the discussion is limited only to the mixed states generated by 
the memory patterns belonging to the first group, 
as mentioned above, the result does not lose generality. 
Hereafter, we discuss the mixed states under this assumption,  
and describe the mixed state as $\mbox{\boldmath{$\gamma$}}^{(s,k)}$. 
Therefore,
among $s$ types of mixed states, 
mixed state $\mbox{\boldmath{$\gamma$}}^{(s,1)}$ 
is considered to be the OR mixed state,
where its $i$th element is given by 
the OR operation through the $i$th elements of the $s$ memory patterns.
The mixed state $\mbox{\boldmath{$\gamma$}}^{(s,s)}$ is 
considered to be the AND mixed state, 
where its $i$th element is given by the AND operation.
Mixed state $\mbox{\boldmath{$\gamma$}} ^{(s,[\frac{s+1}{2}])} $ is 
the majority-decision mixed state that regards the element value 
with more numbers of 0 or 1 for the $i$th elements of $s$ memory patterns.
$[\cdot]$ stands for the Gauss' symbol.
These mixed states can be made to automatically become the equilibrium states 
by choosing an appropriate threshold value,
even if they are under the un-learning condition.

Threshold value $h$ in Eq. (\ref{eq-model1:equilibrium}) is 
determined as follows.
The threshold value is calculated using 
the firing rate of the recalled pattern as described below.
The threshold value obtained by this method 
corresponds approximately with the optimum threshold value 
by which the storage capacity is maximized \cite{Okada1996}.  
Since the firing rate of the equilibrium state is $f$  
when the memory pattern is recalled, 
threshold value $h$ is determined by solving the equation,
\begin{equation} 
 f = \frac{1}{N} \sum_{i=1}^N
  \Theta \left(\sum_{j \neq i}^N J_{ij} x_j^t + h^t \right).
  \label{equationthreshold value}
\end{equation}
When recalling mixed state
$\mbox{\boldmath{$\gamma$}} ^{(s,k)}$,
$f$ in Eq. (\ref{equationthreshold value}) is replaced with
the firing rate $f^{(s,k)}$ 
of mixed state $\mbox{\boldmath{$\gamma$}} ^{(s,k)}$,
\begin{equation} 
 f^{(s,k)} = \mbox{E}[\gamma^{(s,k)}_i]
  = \sum_{v=k}^{s} {}_s \mbox{C}_{v} f^v (1-f)^{s-v},
  \label{equationf_s_k}
\end{equation}
to determine the threshold value.
Where, $\mbox{C}$ is the number of combinations.
The overlap between the state $\mbox{\boldmath{$x$}}$ 
and the memory pattern $\mbox{\boldmath{$\eta$}}^{\mu \nu}$ 
is defined as
\begin{equation}
  m^{\mu \nu} = \frac{1}{Nf(1-f)}
  \sum_{i=1}^N (\eta_i^{\mu \nu}-f) x_i.
  \label {eq:overlap}
\end{equation}
If the state $\mbox{{\boldmath{$x$}}}$ is completely equal
to $\mbox{\boldmath{$\eta$}}^{\mu \nu}$, then $m_\mu=1$.
The overlap between the state {\boldmath{$x$}} 
and the mixed state $\mbox{\boldmath{$\gamma$}} ^{(s,k)}$ 
is defined in a manner similar to Eq. (\ref{eq:overlap}),
\begin{equation}
  M^{(s,k)} = \frac{1}{N f^{(s,k)} (1- f^{(s,k)} )}
  \sum_{i=1}^N (\gamma^{(s,k)}_i - f^{(s,k)}) x_i.
  \label {eq:overlapM}
\end{equation}
If the state $\mbox{{\boldmath{$x$}}}$ is completely equal to
$\mbox{\boldmath{$\gamma$}} ^{(s,k)}$, then $M^{(s,k)}=1$.

%
%
\section{Results}

\subsection{OR mixed state}

In this section,
we explain why 
the storage capacity of the OR mixed state diverges at the sparse limit,
and why the storage capacity of the other mixed states becomes 0. 

We first discuss whether the memory pattern $\mbox{\boldmath $\eta$}^{1,1}$ 
can become the equilibrium state of the present model,
by the following 1step-S/N analysis
\cite{Perez-Vicente1989}\cite{Okada1996}.
If initial state $\mbox{\boldmath $x$}^0$ of the model 
is set as the memory pattern $\mbox{\boldmath $\eta$}^{1,1}$,
internal potential $u_i$ of the $i$th neuron is described as follows 
using Eqs. (\ref{eq-model1:covariance}) and (\ref{eq:overlap}).
\begin{eqnarray}
 u_i &=& \sum_{j \neq i}^N J_{ij} \eta^{1,1}_j + h \nonumber \\
 &=& \sum_{\nu=1}^{s} \sum_{\nu'=1}^{s} (\eta^{1,\nu}_i-f) B_{\nu \nu'} m^{1,\nu'} +h +\bar{z_i},\\
 m^{1,\nu} &=& \frac{1}{Nf(1-f)} \sum_{i=1}^N (\eta^{1,\nu}_i -f) \eta^{1,1}_i, 
	\label{equationS/N1_m} \\
 \bar{z_i} &=& \frac{1}{Nf(1-f)} \sum_{\mu=2}^{\alpha N} \sum_{\nu=1}^{s} \sum_{\nu'=1}^{s}
      \sum_{j \ne i}^{N} (\eta^{\mu \nu}_i-f) B_{\nu \nu'} (\eta^{\mu \nu'}_j-f) \eta^{1,1}_j ,
      \label{equationS/N1_noise}
\end{eqnarray}
where,
at the limit of $N \rightarrow \infty$,
each overlap becomes $m^{1,1}=1,$ $m^{1,2}=,\cdots,=m^{1,s}=0$
by Eqs. (\ref{equationpattern}) and (\ref{eq:overlap}).
The internal potential becomes
\begin{equation}
  u_i = (\eta^{1,1}_i-f) + b\sum_{\nu=2}^{s} (\eta^{1,\nu}_i-f) + h + \bar{z_i}. 
  \label{equationS/N1}
\end{equation}
The first term in this equation is a signal term 
to recall the memory pattern $\mbox{\boldmath $\eta$}^{1,1}$. 
The second term is the
contribution from the other memory patterns belonging to the same group. 
The third term is the threshold value,
which distributes the internal potential to the positive value or negative value. 
And the fourth term $\bar{z}_i$ is the cross-talk noise
that prevents recalling the memory pattern.
According to Eq. (\ref{equationS/N1_noise}),
$\bar{z_i}$ has a normal distribution, 
$N(0, \alpha f s(1+b^2(s-1)))$, at the limit $N \rightarrow \infty$.
When the number of memory patterns is $O(1)$ in $N \rightarrow \infty$,
the loading rate $\alpha$ becomes 0, and the variance of $\bar{z_i}$ becomes 0.

To keep $\eta_i^{1,1}$ as an equilibrium state of the model,
state $x_i$ should be 1 at the next step if the neuron is set to $\eta_i^{1,1}=1$,
and  
state $x_i$ should be 0 at the next step if the neuron is set to $\eta_i^{1,1}=0$.
This method of evaluating the stability of the memory pattern 
using the condition above
is called the 1step-S/N analysis. 
The input (the first and second terms in Eq. (\ref{equationS/N1})) of the neuron 
set to $\eta_i^{1,1}=1$ becomes
\begin{equation}
	(1-f) +b \sum_{\nu=2}^{s} (\eta_i^{1,\nu}-f).
     \label{equationS/N0_Psignal}
\end{equation}
According to this equation,
the input of the neuron that has $\eta_i^{1,2}=,\cdots,=\eta_i^{1,s}=0$
is the smallest, 
and the value is given by
\begin{equation}
	(1-f) -b(s-1)f.
     \label{equationS/N1_Psignal}
\end{equation}
On the other hand,
the input of the neuron set to $\eta_i^{1,1}=0$ becomes 
\begin{equation}
	-f +b \sum_{\nu=2}^{s} (\eta_i^{1,\nu}-f).
     \label{equationS/N0_Msignal}
\end{equation}
According to this equation,
the input of the neuron that has $\eta_i^{1,2}=,\cdots,=\eta_i^{1,s}=1$
is the largest, 
and the value is given by
\begin{equation}
	-f +b(s-1)(1-f).
     \label{equationS/N1_Msignal}
\end{equation}
Let us choose the threshold value 
between Eqs. (\ref{equationS/N1_Psignal}) and (\ref{equationS/N1_Msignal}).
If we assume that the threshold value is the mean value
of Eqs. (\ref{equationS/N1_Psignal}) and (\ref{equationS/N1_Msignal}),
\begin{equation}
     h=-(1-2f)(1+b(s-1))/2,
     \label{threshold-5}
\end{equation}
the memory pattern $\mbox{\boldmath $\eta$}^{1,1}$ can become
the equilibrium state of the model. 
At sparse limit $f \rightarrow 0$, 
the threshold value becomes
\begin{equation}
      h=-(1+b(s-1))/2.
      \label{threshold-6}
\end{equation}

Next, 
we analyze 
whether mixed state $\mbox{\boldmath $\gamma$}^{(s,k)}$ 
can become the equilibrium state of the model, 
and explain why
the OR mixed state $\mbox{\boldmath $\gamma$}^{(s,1)}$ is the only object 
that should be discussed.
If initial state $\mbox{\boldmath $x$}^0$ of the model 
is set as the mixed state $\mbox{\boldmath $\gamma$}^{(s,k)}$,
the internal potential $u_i$ of the $i$th neuron is described as follows 
using Eqs. (\ref{eq-model1:covariance}) and (\ref{eq:overlap}).
\begin{eqnarray}
 u_i &=& \sum_{j \neq i}^N J_{ij} \gamma^{(s,k)}_j + h \nonumber \\
 &=& \sum_{\nu=1}^{s} \sum_{\nu'=1}^{s} (\eta^{1,\nu}_i-f) B_{\nu \nu'} m^{1,\nu'} +h +\bar{z_i},\\
 m^{1,\nu} &=& \frac{1}{Nf(1-f)} \sum_{i=1}^N
  (\eta^{1,\nu}_i -f) \gamma^{(s,k)}_i, 
	\label{equationS/N2_m} \\
 \bar{z_i} &=& \frac{1}{Nf(1-f)} \sum_{\mu=2}^{\alpha N} \sum_{\nu=1}^{s} \sum_{\nu'=1}^{s}
      \sum_{j \ne i}^{N} (\eta^{\mu \nu}_i-f) B_{\nu \nu'} (\eta^{\mu \nu'}_j-f) \gamma^{(s,k)}_j .
      \label{equationS/N2_noise}
\end{eqnarray}
At the limit $N \rightarrow \infty$,
since the overlaps become $m^{1,1}=m^{1,2}=,\cdots,=m^{1,s}$
by Eqs. (\ref{equationpattern}) and (\ref{eq:overlap}),
these overlaps are put with $m^{(s,k)}$
and the internal potential becomes 
\begin{equation}
     u_i = m^{(s,k)} (1+b(s-1))\sum_{\nu=1}^{s} (\eta^{1,\nu}_i-f) +h +\bar{z_i}. 
     \label{equationS/N2}
\end{equation}
The first term in this equation is a signal term 
to recall the mixed state $\mbox{\boldmath $\gamma$}^{(s,k)}$,
the second term is the threshold value,
and the third term $\bar{z}_i$ is cross-talk noise.
Since the overlap $m^{(s,k)}$ determines the strength of the signal term,
let us evaluate its value.
At the sparse limit, 
the overlap $m^{(s,k)}$ is calculated 
based on Eqs. (\ref{equationpattern}) and (\ref{equationS/N2_m})
and becomes 
\begin{equation}
   m^{(s,k)} = \sum_{v=0}^{s-1} 
	    {}_{s-1} \mbox{C}_{v} (1-f)^{s-v-1} f^v  
           \left[ \Theta(1+v-k)-\Theta(v-k) \right] ,
     \label{S/N:m}
\end{equation}
where, $\mbox{C}$ is the number of combinations.
Let us evaluate this equation at the sparse limit.
At sparse limit $f \rightarrow 0$, 
$(1-f)^{s-v-1}$ is 1, and
$f^v$ is 1 in $v=0$ and 0 in $v \ge 1$.
Consequently, the summation for $v$ in the equation
has to be taken into account only when $v=0$.
Therefore, 
the Eq. (\ref{S/N:m}), at the sparse limit, becomes
\begin{equation}
   m^{(s,k)} = \left[ \Theta(1-k)-\Theta(-k) \right].
     \label{S/N:m2}
\end{equation}
This equation says that $m^{(s,k)}=1$ only when $k=1$ (the OR mixed state), 
and that $m^{(s,k)}=0$ when $k \ge 2$ (the other mixed states), 
at the sparse limit.
Note that $\Theta(0) =1$ as in Eq. (\ref{eq-model1:theta}).

\begin{figure}
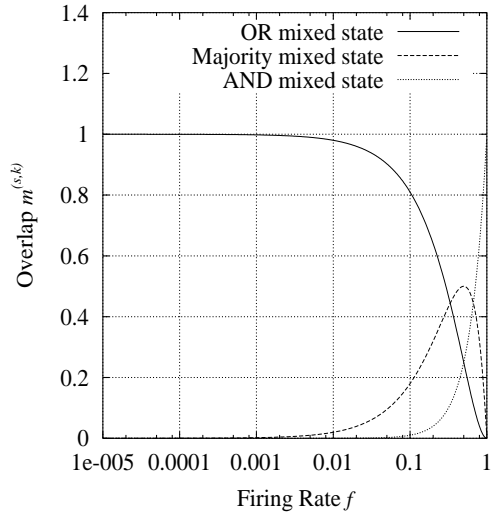

  \caption{Relationship between firing rate $f$ and overlap $m^{(s,k)}$ for $s=3$.}
  \label{fig:1}
\end{figure}

Figure \ref{fig:1} shows the relationship between 
the firing rate $f$ and the overlap $m^{(s,k)}$ for $s=3$.
It shows that $m^{(s,k)} \rightarrow 1$ only when $k=1$ 
and that $m^{(s,k)} \rightarrow 0$ when $k\ge2$,
at the sparse limit.
From these results,
mixed states excluding the OR mixed state do not become the equilibrium state
of the model at the sparse limit.
Therefore,
 we limit our remaining discussion to the OR mixed state. 

To keep the OR mixed state $\gamma_i^{(s,1)}$ as an equilibrium state of the model,
state $x_i$ should be 1 at the next step if the neuron is set to $\gamma_i^{(s,1)}=1$,
and  
state $x_i$ should be 0 at the next step if the neuron is set to $\gamma_i^{(s,1)}=0$.
If the neuron is set to $\gamma_i^{(s,1)}=1$,
since one or more of the $s$ memory patterns included in the signal term is $1$, 
the input (the first term in Eq. (\ref{equationS/N2})) of the neuron becomes
\begin{equation}
	m^{(s,1)}(1+b(s-1))(n-sf),\quad (1 \le n \le s).
\end{equation}
According to this equation,
the input of the neuron with $n=1$ is the smallest, 
and the value is given by
\begin{equation}
	m^{(s,1)}(1+b(s-1))(1-sf).
     \label{equationS/N2_Psignal}
\end{equation}
In the neuron set to $\gamma_i^{(s,1)}=0$,
since all of the memory patterns included in the signal term are $0$, 
the input of the neuron is
\begin{equation}
	m^{(s,1)}(1+b(s-1))(-sf).
     \label{equationS/N2_Msignal}
\end{equation}
Let us choose the threshold value 
between Eqs. (\ref{equationS/N2_Psignal}) and (\ref{equationS/N2_Msignal}). 
If we assume the threshold value is the mean value
of Eqs. (\ref{equationS/N2_Psignal}) and (\ref{equationS/N2_Msignal}),
\begin{equation}
      h=-m^{(s,1)}(1+b(s-1))(1-2sf)/2,
     \label{threshold-7}
\end{equation}
The OR mixed state $\mbox{\boldmath $\gamma$}^{(s,1)}$ can become 
the equilibrium state of the model. 
At sparse limit $f \rightarrow 0$, the threshold value reaches 
the following value because $m^{(s,k)}=1$,
\begin{equation}
      h=-(1+b(s-1))/2.
     \label{threshold-8}
\end{equation}
Since this threshold value and that of the memory pattern 
in Eq. (\ref{threshold-6}) are the same forms,
those values are the same even if the b and/or s change.

From these results,
we found that
only the memory pattern and the OR mixed state become the equilibrium states at the sparse limit,
and that the memory pattern and the OR mixed state can be made to easily coexist 
at the sparse limit because their threshold values are the same.
We proved that the properties of the present model does not change from
those of the sparsely encoded conventional model (Kimoto \& Okada, 2001).

%
%
\subsection{Quantitative evaluation using SCSNA analysis}

Using the 1step-S/N analysis, 
we proved that the memory pattern and the OR mixed state are 
the objects that should be evaluated. 
However,
the cross-talk noise in Eqs. (\ref{equationS/N1_noise}) and (\ref{equationS/N2_noise})
cannot be quantitatively evaluated using the 1step-S/N analysis.
In this section,
we quantitatively analyze the storage capacities and the threshold values 
of the memory pattern and of the OR mixed state 
using the SCSNA(self-consistent signal-to-noise analysis) 
proposed by Shiino and Fukai (1992).
The order parameter equations of the equilibrium state derived by the SCSNA
reach the order parameter equations of the equilibrium state
derived by the replica theory of the statistical mechanical theory.

We consider the case where the equilibrium state $\mbox{\boldmath $x$}$
has nonzero overlaps, 
$m^{1, \nu} = \frac{1}{Nf(1-f)} \sum_{i=1}^{N} (\eta^{1, \nu}_i-f) x_i$,
with $s$ memory patterns $\mbox{\boldmath $\eta$}^{1,\nu}, (1 \le \nu \le s)$.
This means the case 
where the memory pattern belonging to the first group 
or the mixed state generated by the memory patterns of the first group 
is recalled. 
To derive the SCSNA order parameter equations of the present model,
we first transform $J_{ij}$ into the following form, 
which can easily be applied to the SCSNA.
The synaptic coupling $J_{ij}$ is divided into 
the term related to the first group memory pattern
and the other term:
\begin{eqnarray}
   J_{ij} &=& \frac {1}{Nf(1-f)} \sum_{\nu=1}^{s} \sum_{\nu'=1}^{s}
	      (\eta^{1,\nu}_i-f) B_{\nu \nu'} (\eta^{1,\nu'}_j-f)      \nonumber \\
          & & + \frac {1}{Nf(1-f)} \sum_{\mu=2}^{\alpha N}
              \sum_{\nu=1}^{s} \sum_{\nu'=1}^{s}
	      (\eta_i^{\mu \nu} - f) B_{\nu \nu'} (\eta_j^{\mu \nu'} - f).
     \label{eq-model1:covariance2}
\end{eqnarray}
We introduce a set of i.i.d. patterns $\sigma^{\mu \nu}_i$ 
to the $(\eta^{\mu \nu}_i-f)$:
\begin{eqnarray}
     \mbox{E}[\sigma^{\mu \nu}_i] &=& 0, \\
     \mbox{E}[\sigma^{\mu \nu}_i \sigma^{\mu \nu'}_i] &=& 0. \quad (\nu \ne \nu')
     \label{eq-model1:E_sigma}
\end{eqnarray}
The second term of the synaptic coupling in Eq. (\ref{eq-model1:covariance2}) is rewritten using $\sigma^{\mu \nu}_i$ as
\begin{eqnarray}
   J_{ij} &=& \frac {1}{Nf(1-f)} \sum_{\nu=1}^{s} \sum_{\nu'=1}^{s}
	      (\eta^{1,\nu}_i-f) B_{\nu \nu'} (\eta^{1,\nu'}_j-f)      \nonumber \\
          & & + \frac {1}{Nf(1-f)} \sum_{\mu=2}^{\alpha N}
              \sum_{\nu=1}^{s} \sum_{\nu'=1}^{s}
	      \sigma^{\mu \nu}_i B_{\nu \nu'} \sigma^{\mu \nu'}_j.
     \label{eq-model1:covariance3}
\end{eqnarray}
Let ${\bf e}^\nu$,$(\nu=1,2,\cdots,s)$ be a set of 
$s$ dimensional normalized eigenvectors of the matrix ${\mbox{\bf B}}$ 
in Eq. (\ref{eq-model1:cross-colleration}).
We introduce a set of rotated patterns, 
$\bar{\mbox{\boldmath{$\sigma$}}}^\mu_i=(\bar{\sigma}^{\mu,1}_i,\bar{\sigma}^{\mu,2}_i,\cdots,\bar{\sigma}^{\mu,s}_i)$:
\begin{eqnarray}
	\mbox{\boldmath{$\sigma$}}^\mu_i &=& {\bf T} \bar{\mbox{\boldmath{$\sigma$}}}^\mu_i,
     \label{eq-model1:rotated_pattern} \\
	{\bf T} &=& ({\bf e}^1,{\bf e}^2,\cdots,{\bf e}^{s}).
     \label {eq-model1:eigenvectors} 
\end{eqnarray}
Since the rotated patterns $\bar{\mbox{\boldmath{$\sigma$}}}^\mu_i$ 
is simply rotated using the matrix ${\mbox{\bf T}}$,
the distribution is the same as that of $\mbox{\boldmath{$\sigma$}}^\mu_i$.
Using rotated pattern $\mbox{\boldmath{$\sigma$}}^\mu_i$,
the synaptic coupling $J_{ij}$ in Eq. (\ref{eq-model1:covariance3}) 
is rewritten as
\begin{eqnarray}
   J_{ij} &=& \frac {1}{Nf(1-f)} \sum_{\nu=1}^{s} \sum_{\nu'=1}^{s}
	      (\eta^{1,\nu}_i-f) B_{\nu \nu'} (\eta^{1,\nu'}_j-f)
                                                     \nonumber \\
          & & + \frac {1}{Nf(1-f)} \sum_{\mu=2}^{\alpha N}
              \sum_{\nu=1}^{s} 
	      \lambda^\nu \bar{\sigma}^{\mu \nu}_i \bar{\sigma}^{\mu \nu}_j,
     \label{eq-model1:covariance4}
\end{eqnarray}
where
$\lambda^\nu$ is the eigenvalue of the matrix ${\mbox{\bf B}}$ for normalized eigenvector ${\bf e}^\nu$:
$\lambda^1=1+(s-1)b^2$C
$\lambda^\nu=1-b^2,(2 \le \nu \le s)$.
The internal potential $u_i$ is written as follows,
using $J_{ij}$ in Eq. (\ref{eq-model1:covariance4}) 
and the overlap $m^{1,\nu}$ in Eq. (\ref{eq:overlap}):
\begin{eqnarray}
   u_i &=& \sum_{j \ne i}^N J_{ij} x_j \nonumber \\
       &=& \sum_{\nu=1}^{s} \sum_{\nu'=1}^{s} (\eta^{1,\nu}_i-f) B_{\nu \nu'} m^{1,\nu'}
           +\sum_{\mu=2}^{\alpha N} \sum_{\nu=1}^{s} 
	      \lambda^\nu \bar{\sigma}^{\mu \nu}_i \bar{m}^{\mu \nu}
           - \alpha s x_i,
\end{eqnarray}
where,
\begin{equation}
    \bar{m}^{\mu \nu} = \frac{1}{Nf(1-f)} 
          \sum_{i=1}^{N} \bar{\sigma}^{\mu \nu}_i x_i.
\end{equation}

Hereafter,
following the SCSNA \cite{Shiino1992}\cite{Fukai1999}\cite{Toya2000},
we can easily derive the SCSNA order parameter equations:
\begin{eqnarray}
  Y_i
   &=& \Theta \left(\sum_{\nu=1}^{s} \sum_{\nu'=1}^{s} (\eta^{1,\nu}_i-f) B_{\nu \nu'} m^{1,\nu'}
	  + \Gamma Y_i + \sqrt{\alpha r} z +h^t
	  \right), \label{eq-model1:Y2} \\
 m^{1,\nu} &=& 
	\frac{
  \int_{-\infty}^{\infty} D_z <(\eta^{1,\nu}_i-f) Y_i>_{\mbox{\boldmath $\eta$}}
	}
	{f(1-f)}
  , \label{eq-model1:m2} \\
 q &=& \int_{-\infty}^{\infty} D_z <(Y_i)^2>
  _{\mbox{\boldmath $\eta$}}, \label {eq-model1:q2} \\
 U &=& \frac{1}{\sqrt{\alpha r}}
  \int_{-\infty}^{\infty} D_z z <Y_i>_{\mbox{\boldmath $\eta$}},
  \label {eq-model1:U2}
  \\
 D_z &=& \frac{dz}{\sqrt{2 \pi}} \exp(-\frac{z^2}{2}),
  \label {eq-model1:D_z2}
  \\
 r &=& q \sum_{\nu=1}^{s} \frac{(\lambda^\nu)^2}{(1-\lambda^\nu U)^2},
  \label {eq-model1:r2} \\
 \Gamma &=& \alpha \sum_{\nu=1}^{s} 
     \frac{ (\lambda^\nu)^2 U}{1-\lambda^\nu U},
  \label{eq-model1:Gamma2}
\end{eqnarray}
where,
$< \cdots >_{\mbox{\boldmath $\eta$}}$ stands for an ensemble average 
over the first group memory patterns, 
$\mbox{\boldmath $\eta$}=({\bf \eta}_i^{1,1}, {\bf \eta}_i^{1,2}, \cdots, {\bf \eta}_i^{1,s})$.
Equation (\ref{eq-model1:Y2}) may have more than one solution 
by the ``Maxell equal area rule'',
which originated in thermodynamics.
According to this rule,
$Y_i$ in Eq. (\ref{eq-model1:Y2}) can be rewritten as
\begin{equation}
  Y_i = \Theta \left(\sum_{\nu=1}^{s} \sum_{\nu'=1}^{s} (\eta^{1,\nu}_i-f) B_{\nu \nu'} m^{1,\nu'}
	  + \frac{1}{2}\Gamma  + \sqrt{\alpha r} z +h^t
	  \right).
	\label{eq-model1:Y3} \\
\end{equation}
We can continuously
 obtain the following SCSNA order parameter equations
by integrating Eqs. (\ref{eq-model1:m2})-(\ref{eq-model1:Y3}):
\begin{eqnarray}
  m^{1,\nu} &=&	< 
		\frac{\eta^{1,\nu}-f}{2f(1-f)} 
	  	\mbox{erf}(
                      \frac{ \sum_{\nu=1}^s \sum_{\nu'=1}^s (\eta^{1,\nu}-f) B_{\nu \nu'} m^{1,\nu'}
                      +h^t +\frac{\Gamma}{2}}
		{\sqrt{2 \alpha r}}
	)>_{\mbox{\boldmath $\eta$}}, \nonumber \\
    && \qquad\qquad\qquad\qquad\qquad\qquad\qquad\qquad\qquad\qquad (1 \le \nu \le s), 
 \label{eq-model1:m3} \\
 q &=& \frac{1}{2} + \frac{1}{2} <
		\mbox{erf}(\frac{
                    \sum_{\nu=1}^s \sum_{\nu'=1}^s (\eta^{1,\nu}-f) B_{\nu \nu'} m^{1,\nu'}
                    +h^t +\frac{\Gamma}{2}}
		{\sqrt{2 \alpha r}}
	>_{\mbox{\boldmath $\eta$}}, 
 \label{eq-model1:q3} \\
 U &=& \frac{1}{\sqrt{2\pi \alpha r}}< 
		\mbox{exp}(-(\frac{
                       \sum_{\nu=1}^s \sum_{\nu'=1}^s (\eta^{1,\nu}-f) B_{\nu \nu'} m^{1,\nu'}
                       +h^t +\frac{\Gamma}{2}}
		{\sqrt{2 \alpha r}})^2
	>_{\mbox{\boldmath $\eta$}},\nonumber \\
 &&\label{eq-model1:U3} \\
 r &=& q \sum_{\nu=1}^{s} \frac{(\lambda^\nu)^2}{(1-\lambda^\nu U)^2},
  \label {eq-model1:r3} \\
 \Gamma &=& \alpha \sum_{\nu=1}^{s} \frac{(\lambda^\nu)^2 U}{1-\lambda^\nu U}.
  \label{eq-model1:Gamma3}
\end{eqnarray}
We can derive the relationship between $m^{1,\nu}$ and $\alpha$
by solving the simultaneous equations (\ref{eq-model1:m3})-(\ref{eq-model1:Gamma3}).
Note that threshold value $h$ must be determined
to solve these equations (\ref{eq-model1:m3})-(\ref{eq-model1:Gamma3})
because they include $h$. 
The following order parameter equation,
which determines the threshold value,
is derived from Eq. (\ref{equationthreshold value}):
\begin{eqnarray}
    f &=& \int_{-\infty}^{\infty} D_z <Y_i>_{\mbox{\boldmath $\eta$}}, \nonumber \\
      &=& \frac{1}{2} + \frac{1}{2} <
		\mbox{erf}(\frac{
                    \sum_{\nu=1}^s \sum_{\nu'=1}^s (\eta^{1,\nu}-f) B_{\nu \nu'} m^{1,\nu'}
                    +h^t +\frac{\Gamma}{2}} {\sqrt{2 \alpha r}})
	>_{\mbox{\boldmath $\eta$}}. 
        \label {eq-model1:threshold2}
\end{eqnarray}
In the case of recalling the OR mixed state $\mbox{\boldmath{$\gamma$}} ^{(s,1)}$,
the threshold value is determined by 
the following order parameter equation 
in which $f$ in Eq. (\ref{eq-model1:threshold2}) is replaced with
the firing rate $f^{(s,1)}$ of the OR mixed state:
\begin{equation}
    f^{(s,1)} = \frac{1}{2} + \frac{1}{2} <
		\mbox{erf}(\frac{
                    \sum_{\nu=1}^s \sum_{\nu'=1}^s (\eta^{1,\nu}-f) B_{\nu \nu'} m^{1,\nu'}
                    +h^t +\frac{\Gamma}{2}} {\sqrt{2 \alpha r}})
	>_{\mbox{\boldmath $\eta$}}. 
        \label {eq-model1:threshold3}
\end{equation}
The overlap $M^{(s,1)}$ between the equilibrium state $\mbox{\boldmath $x$}$
and the OR mixed state $\mbox{\boldmath $\gamma$}^{(s,1)}$ 
is defined by
\begin{equation}
  M^{(s,1)} = 
	< \frac{\gamma^{(s,1)}-f^{(s,1)}} {2f^{(s,1)}(1-f^{(s,1)})}
  	  	\mbox{erf}(\frac{ 
                         \sum_{\nu=1}^s \sum_{\nu'=1}^s (\eta^{1,\nu}-f) B_{\nu \nu'} m^{1,\nu'}
			 +h^t +\frac{\Gamma}{2}}
		{\sqrt{2 \alpha r}}
	)>_{\mbox{\boldmath $\eta$}}. 
  \label {eq-model1:M}
\end{equation}

\begin{figure}
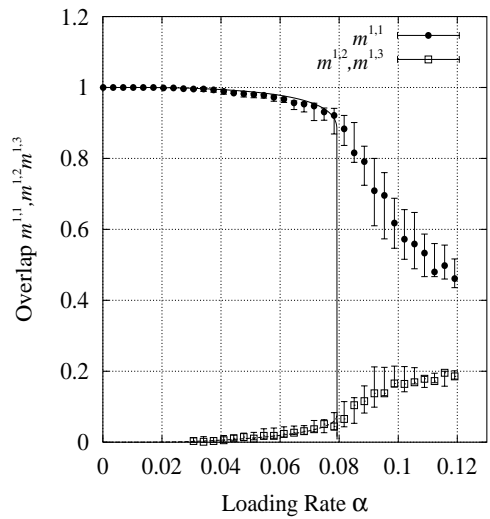

  \caption{
    Overlap $m^{1,1}, m^{1,2},$ and $m^{1,3}$ for various loading rates $\alpha$
    when recalling memory pattern $\mbox{\boldmath $\eta$}^{1,1}$;
    $f=0.1, b=0.25, s=3, N=10,000$.}
  \label{fig:2}
\end{figure}

Next,
to investigate the effectiveness of the SCSNA to the present model, 
we compare the results of the SCSNA with those of the computer simulation. 
Figure \ref{fig:2} shows 
the overlaps $m^{1,1}, m^{1,2},$ and $m^{1,3}$ for various loading rates $\alpha$, 
when recalling the memory pattern $\mbox{\boldmath $\eta$}^{1,1}$.
We used $s=3$, $b=0.25$, and $f$=0.1.
The data points and error bars show the results of the computer simulation,
and the lines connecting the data points are the results of the SCSNA.
In the computer simulation,
number of neurons $N$ was set to 10,000,
and the simulation was run 11 times for each parameter.
The data points show the median, 
and the ends of the error bars show the $1/4$ 
and $3/4$ deviations.
The results of the SCSNA show that
the overlap $m^{1,1}$ gradually decreased from 1
as the loading rate $\alpha$ was increased from 0,
and that
the equilibrium state disappeared at $\alpha \simeq 0.08$. 
The storage capacity $\alpha_c$ was then $\alpha_c \simeq 0.08$.
The results of the computer simulation corresponded to
those of the SCSNA when $0 \le \alpha \le 0.08$, and
the equilibrium state became unstable at $\alpha > 0.08$.
The results of the SCSNA and the simulation also corresponded well
for the overlaps $m^{1,2}$ and $m^{1,3}$.

\begin{figure}
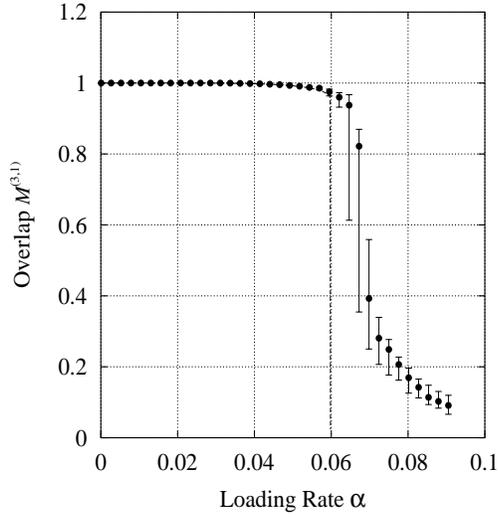

  \caption{Overlap $M^{(3,1)}$ for various loading rates $\alpha$
    when recalling OR mixed state $\mbox{\boldmath{$\gamma$}}^{(3,1)}$;
    $f=0.1, b=0.25, s=3, N=10,000$.}
  \label{fig:3}
\end{figure}

Figure \ref{fig:3} shows
the overlap $M^{(3,1)}$ for various loading rates
when recalling the OR mixed state $\mbox{\boldmath $\gamma$}^{(3,1)}$.
We used $s=3$, $b=0.25$, and $f$=0.1.
The data points and error bars show the results of the computer simulation,
and the line connecting the data points shows the result of the SCSNA.
The results of both corresponded well.
Since 
the results of the SCSNA explain the results of simulation fairly well,
as shown in Figs. \ref{fig:2} and \ref{fig:3},
we hereafter examine
the properties of the memory pattern and of the OR mixed state 
using the SCSNA. 

\begin{figure}
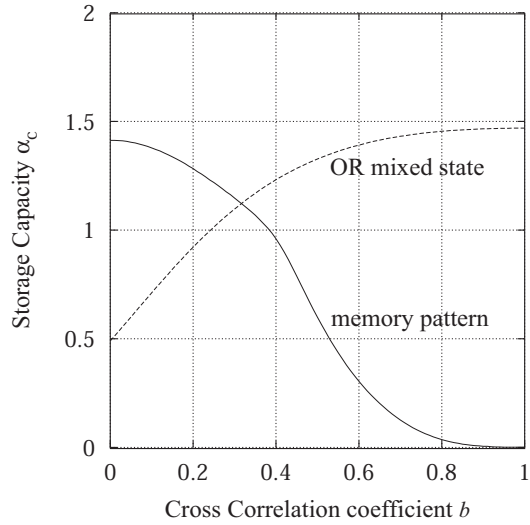

  \caption{Storage capacity $\alpha_c$ of memory pattern and OR mixed state 
    for various $b$; $f=0.01, s=3$.}
  \label{fig:4}
\end{figure}

Figure \ref{fig:4} shows 
the storage capacity of the memory pattern and of the OR mixed state
for various value $b$.
We used $s=3$ and $f$=0.01.
The storage capacity $\alpha_c$ of the memory pattern
was about $1.4$ at $b=0$. 
Since each group includes $s$ memory patterns, 
the total number of memory patterns that can become an equilibrium state 
is $s \alpha_c N \simeq 4.2 N$. 
Since $b=0$ means the model without the cross-correlation term,
the storage capacity is equal that
of the conventional model
which sparsely encoded by the firing rate $f=0.01$.
The storage capacity of the memory pattern gradually decreased
as $b$ was increased. 
The storage capacity of the OR mixed state was about $0.5$ at $b=0$, 
gradually increased as $b$ was increased,
and  
was about $1.45$ at $b=1.0$. 
Thus, 
the $b$ dependency of the storage capacity of the memory pattern
is different from that of the OR mixed state. 

\begin{figure}
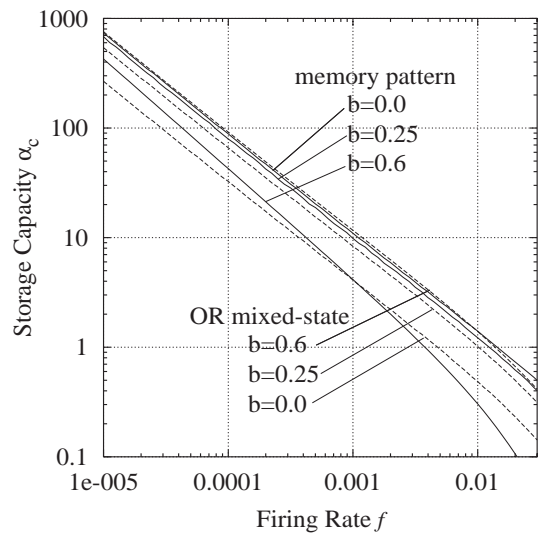

  \caption{Asymptotic properties of storage capacity $\alpha_c$ 
    for memory pattern and OR mixed state; $b=0, 0.25, 0.6$, $s=3$.}
  \label{fig:5}
\end{figure}

To analyze the properties of the memory pattern and of the OR mixed state at the sparse limit 
$f \rightarrow 0$,
we plotted the asymptotes of the storage capacity in Fig. \ref{fig:5}.
We used $b=0, 0.25, 0.6$.
Since $b=0$ means the model without the cross-correlation term,
the present model corresponds to the conventional model.
The asymptotes of the memory pattern and of the OR mixed state
diverged as $1/|f \log f|$ at the sparse limit,
as reported by Kimoto and Okada (2001).
Even when $b$ was 0.25 or 0.6,
at sparse limit $f \rightarrow 0$,
the asymptotes of the memory pattern and of the OR mixed state 
diverged as $1/|f \log f|$.
Thus,
the asymptotic properties of the storage capacity of the present model
are the same as those of the conventional model.
Moreover,
the storage capacity of the memory pattern decreased as $b$ was increased,
and the storage capacity of the OR mixed state increased as $b$ was increased. 
These results correspond to those shown in Fig. \ref{fig:4}.

Figure \ref{fig:6} shows the threshold values
calculated using Eqs. (\ref{eq-model1:threshold2}) and (\ref{eq-model1:threshold3})
to recall the memory pattern and the OR mixed state.
We used $f=0.01, s=3, b=0, 0.25, 0.6$.
The lines are drawn to the point where the loading rates reach the storage capacities. 
As shown in the figure, when the loading rate was small,
the threshold values needed to recall the memory pattern were
nearly equal to the threshold values needed to recall the OR mixed state.
When the loading rate was increased, 
the threshold values at $b=0$ approximately corresponded,
while at $b=0.25$ or $0.6$ they gradually separated. 
Although 
the threshold value of the memory pattern corresponded to 
that of the OR mixed state in the 1step-S/N analysis,
they did not necessarily correspond in the quantitative analysis using SCSNA.
In the conventional model, 
the threshold value needed to recall the memory pattern approximately corresponds to the threshold value needed to recall the mixed state 
at the sparse limit (Kimoto \& Okada, 2001).
We found that 
the properties of the threshold value of the present model
differ from those of the conventional model.

Here,
we examine 
whether the results obtained by the 1step-S/N analysis are much different from 
the properties of the threshold value of the model.
Consider,
a slight change in the threshold value set in the model 
from that shown in Fig.\ref{fig:6}. 
The equilibrium state of the model will shift to 
the neighborhood of the equilibrium state from the memory pattern or from the OR mixed state.
Moreover,
when the threshold value exceeds the boundary value,
a first order phase transition will occur,
and the equilibrium state of the model will rapidly change to a state 
having zero overlap with the memory pattern or with the OR mixed state.
Figure \ref{fig:7} shows 
the regions of the threshold values
in which the equilibrium state of the model 
stayed in the neighborhood of the memory pattern or of the OR mixed state. 
We used $f=0.01, s=3, b=0, 0.25, 0.6$.
These regions narrowed gradually 
as the loading rate was increased,
and the regions disappeared when the loading rate reached the storage capacity. 
Let's see the regions 
until either the loading rate of the memory pattern or of the OR mixed state reached the storage capacity.
When $b=0$ or $0.6$, the regions overlapped each other,
meaning that the memory pattern and the OR mixed state can coexist at the same threshold value.
When $b=0.25$, 
the region overlapped each other until the loading rate $\alpha$ reached $0.6$. 
Thus,
the memory pattern and the OR mixed state can become 
the equilibrium state of the model using the same threshold value,
if the equilibrium state can be expanded to the neighborhood 
of the memory pattern or of the OR mixed state. 
These results confirm that
the results of the 1step-S/N analysis qualitatively indicate the properties of the model.

%
%
\section{Conclusion}

We investigated the properties of the mixed states 
in an associative memory model 
that includes cross-correlation terms in the learning method.

We found that,
under the condition of a fixed firing rate,
the storage capacities of the memory pattern decreases and that of the OR mixed state increases,
as the strengths of the cross-correlation terms are increased.
We also found that, at the sparse limit,
the storage capacity 
of the memory pattern and of the OR mixed state diverges,
and that of the other mixed states becomes 0. 
Additionally, we found that
the threshold value needed to recall the memory pattern does not correspond to 
the threshold value needed to recall the mixed state
in the present model.
However, 
if the equilibrium state can be expanded
to the neighborhood of the memory pattern and the OR mixed state, 
both thresholds can be set to the same value. 

Of further interest,
although the present model, that of Parga and Rolls, and that of Elliffe et al.
use uniformly distributed patterns (uncorrelated patterns),
in general, 
the views of the same object do not appear to be mutually uncorrelated patterns. 
We will discuss about the associative memory model storing correlated patterns elsewhere.

%
%

%
%
\newpage

\newpage

\begin{figure}
  \caption{Threshold values needed to recall memory pattern and OR mixed state;
    $f=0.01, s=3, b=0, 0.25, 0.6$.}
 \label{fig:6}
\end{figure}

\begin{figure}
  \caption{Threshold values in which equilibrium state stay 
    in neighborhood of memory pattern and OR mixed state;
        $f=0.01, s=3, b=0, 0.25, 0.6$.}
  \label{fig:7}
\end{figure}
\end{document}